\documentclass[pra,a4paper,nofootinbib,twocolumn,showpacs,preprintnumbers,amsmath,amssymb,floatfix,amstex]{revtex4}
\newcommand{\ket}[1]{|#1\rangle}                	%
\newcommand{\bra}[1]{\langle #1}               		%
\usepackage{graphicx,graphics,wrapfig,rotating}     	% Include figure files
\usepackage{dcolumn}                    		% Align table columns on decimal point
\usepackage{bbm, bm,fancybox}                		% bold math
\usepackage{times,euscript,eufrak,oldgerm}              %
\usepackage[german,english]{babel}			%
\usepackage{psfrag}
\usepackage{color}

\begin{document}
\title{Measuring multipartite concurrence with a single factorizable observable} %
\author{Leandro Aolita}
\email{aolita@if.ufrj.br}
\affiliation{%
Instituto de F\'\i sica, Universidade Federal do Rio de Janeiro. Caixa Postal
68528, 21941-972 Rio de Janeiro, RJ, Brasil\\
}
\author{Florian Mintert}
\affiliation{%
Department of Physics, Harvard University,
17 Oxford Street, Cambridge Massachusetts, USA}

\date{\today}%
\begin{abstract}
We show that, for any composite system with an arbitrary number of
finite--dimensional subsystems, it is possible to directly measure the multipartite concurrence of pure
states by detecting only one single factorizable observable, provided that two copies of the composite state are available.
This result can be immediately put into practice  in trapped--ion and entangled--photon experiments.
\end{abstract}
\maketitle

%%%%%%%%%%%%%%%%%%%%%%%%%%%%%%%%%%%%%%%%%%%%%%%%%%%%%%%%%%%%%%%%%%%%%%%%%%%%%%%%%%%%%%%%%%%%%%%%%%%%%%%%%%%%%%%%%%%%%%%%%%%%%%%%
\emph{Introduction.}--- A number of measures have been proposed to quantify entanglement (see \cite{Plenio} and references therein).
Originally defined as an auxiliary quantity for the algebraic evaluation of entanglement of formation of two--qubit
systems,
concurrence \cite{Hills-Wooters} is an entanglement measure in its own right \cite{Wooters}.
For the two--qubit case it
has a one--to--one correspondence with entanglement of formation \cite{Wooters}, and it can be generalized to arbitrary--dimensional bipartite \cite{Rungta, Ulhmann} and multipartite
\cite{Meyer-Wallach-Brennen, Andre} systems. Moreover, for pure states it can be interpreted as the expectation value of a Hermitian operator, and thus it can be
measured, if two copies of the state are available \cite{Flo, Flo-Review}.
As a matter of fact, it was this reinterpretation of concurrence in terms of copies of the state that led
to the first direct experimental observation of an entanglement measure \cite{Steve}.
There, a two--qubit entangled state and its copy were encoded in the polarization and transverse momentum degrees of freedom,
respectively, of two twin photons generated via parametric
down conversion; and concurrence was measured by detecting only a single two--qubit joint probability.

\par On the other hand, the experimental progress seen in the last few years in the production and coherent manipulation of
multiparticle entangled states is tremendous. Three photon W--type entanglement has been observed \cite{Eibl, Witnesses}; and three
\cite{Bowmeester, Pan}, four \cite{Pan2, Witnesses} and five \cite{Zhao} photon Greenberger--Horn--Zeilinger(GHZ) entangled states are now realizable. A two--atom--one--photon
GHZ state has been experimentally demonstrated \cite{Raschenbeutel}; three \cite{Roos}, four \cite{Sackett}  and up to six
\cite{Leibfried} ion GHZ states have also been reported; and, very recently, a technique for scalable and deterministic
production of W--type entangled states has successfully generated entangled W--states of up to eight ions \cite{Haeffner}.
Nevertheless, there exists a big mismatch between the progress made on the production and manipulation of multiparticle entanglement
and its experimental quantification. In the  experiments just mentioned, multipartite entanglement was verified either through the use of quantum state tomography
\cite{Roos, Haeffner}, quantum non--locality tests \cite{Eibl, Bowmeester, Pan, Pan2, Zhao, Raschenbeutel}, or entanglement
witnesses \cite{Sackett, Roos, Leibfried, Witnesses, Haeffner}. Quatum state tomography \cite{White, Roos2} provides a complete description of the state, though is very
disadvantageous from the point of view of scalability. Quantum non--locality tests
and entanglement witnesses \cite{HoroWitness}, in contrast, require the measurement of only a few observables; but each of them allows to detect the entanglement of only a small class of states,
so that -- in practical terms -- some a priori knowledge of the state is necessary.
%In \cite{Steve}, the concurrence of a two--qubit pure sate was measured through one single observable. In \cite{Leibfried}, entanglement was detected through the use of
%an entanglement witness capable to detect GHZ--type entanglement. In all the other experiments cited above, though, entanglement
%was detected through  quantum state tomography \cite{White, Roos2}, which provides a complete description of the quantum correlations of the state, but requires the full
%reconstruction of the density operator. Also, remarkable experiments detecting three and four photon entanglement through
%the use of quantum non--locality tests \cite{Pan} and entanglement witnesses \cite{Witnesses} have been carried out.
%Quantum non--locality tests and entanglement witnesses, in turn, allow to detect the entanglement of only a small class of states,
%so that -- in practical terms -- some a priori knowledge of the state is necessary.
%\fi
And, moreover, they typically provide a qualitative description, but do not define an entanglement measure.
Therefore a simple scheme -- involving as few measurements as possible -- to experimentally measure entanglement of multipartite systems is highly desirable.

\par In this paper we show that  for a composite system with an arbitrary number of
finite--dimensional subsystems it is possible to directly measure the multipartite concurrence of pure
states by detecting only one single factorizable observable, provided that two copies of the composite state are available.
This allows for a generalization of the single--setting measurement scheme used in \cite{Steve} to arbitrary dimensional multipartite systems.
In particular, the scheme is directly applicable to trapped--ion and entangled--photon experiments,  which we also discuss.

\emph{Representation of concurrence using two copies of the state}.---
It was shown in \cite{Flo-Review} that the
concurrence $C_{N}$ of an $N$--partite--system pure state $\ket{\Psi_{N}}\in{\cal H}$,
can be expressed as the following expectation value with respect to two copies of $\ket{\Psi_{N}}$:
\begin{equation}
\label{Product}
C_{N}(\Psi_{N})=\sqrt{\bra{\Psi_{N}}|\otimes\bra{\Psi_{N}}|A\ket{\Psi_{N}}\otimes\ket{\Psi_{N}}}\ .
\end{equation}
Here $A$ is a Hermitian operator acting on ${\cal H}\otimes{\cal H}$, {\it i. e.}, on
${\cal H}_{1}\otimes\hdots\otimes{\cal H}_{N}\otimes{\cal H}_{1}\otimes\hdots\otimes{\cal H}_{N}$, where ${\cal H}_{i}$,
with $1\le i \le N$,  is the Hilbert space associated to the $i^ {th}$ subsystem, in terms of which the composite system Hilbert
space ${\cal H}$ factorizes.
The operator $A$ can be written as:
\begin{equation}
\label{Projectors}
A=4\sum_{\{s_{j_{i}}=\pm\}^{+}}P^{1}_{s_{1_{i}}}\otimes\hdots\otimes P^{N}_{s_{N_{i}}},
\end{equation}
where $P^{j}_{+}$, and $P^{j}_{-}$, $(1\le j\le N)$ are the projectors onto the symmetric and antisymmetric subspaces
${\cal H}_{j}\odot{\cal H}_{j}$, and ${\cal H}_{j}\wedge{\cal H}_{j}$, respectively, of the Hilbert space
${\cal H}_{j}\otimes{\cal H}_{j}$ that describes the two copies of the $j^{th}$ subsystem.
The antisymmetric subspace is spanned by all states that acquire a phase shift of $\pi$ upon the exchange of the two involved copies,
whereas the symmetric subspace is spanned by those states that acquire no phase shift at all.
The summation is
restricted to the set $\{s_{j_{i}}=\pm\}^{+}$ composed of all possible ways of sorting the symbols `$+$' and `$-$' in an
$N$--long string, such that the total amount of `$-$' symbols is an even number, and excluding the completely symmetric case with no `$-$' symbols at all, {\it i.e.}
$s_{1_{i}} s_{2_{i}} ...  s_{N_{i}}=++ ... +$.
The prefactor $4$ in Eq.~(\ref{Projectors}) is only a normalization factor so that $C_{N}$ reduces to the original concurrence \cite{Wooters} for the two--qubit case.
When expressed in terms of the reduced density matrices $\varrho_{i}$, $C_{N}$, as given by Eq. (\ref{Product}),
coincides with the multipartite concurrence introduced in \cite{Andre}:
\begin{equation}
\label{ConcAndre}
C_{N}(\Psi_{N})=2^{1-N/2}\sqrt{(2^{N}-2)-\sum_{i}\mbox{Tr}\varrho_{i}^{2}}\ ,
\end{equation}
where the index $i$ labels all $(2^{N}-2)$ subsets of the $N$--particle system;
and the $\varrho_{i}$ are the reduced density matrices of all $1$ to $N-1$ partite subsystems \cite{Flo-Review}.
$C_{N}$ vanishes exactly for $N$--separable states and
allows for a meaningful comparison of entanglement between systems with different numbers of subsystems,
since the $N$--partite concurrence $C_{N}(\Psi_{N})$ reduces to the $(N-1)$--partite concurrence
$C_{N}(\Psi_{N})=C_{N-1}(\Psi_{N-1})$
for any state $\ket{\Psi_{N}}=\ket{\Psi_{N-1}}\otimes\ket{\phi}$
that factorizes into an $(N-1)$--partite state and a one--partite remainder.
Finally, for  $N=2$,  Eq. (\ref{ConcAndre}) yields the arbitrary--dimensional bipartite concurrence defined in \cite{Rungta}.

\emph{$C_{N}$ in terms of a single factorizable observable}.---
Any term $P^{1}_{s_{1_{i}}}\otimes\hdots\otimes P^{N}_{s_{N_{i}}}$
with an even number of antisymmetric factors,
projects onto states that are globally symmetric,
{\it i.e.} that are symmetric with respect to the exchange of the two copies of the entire system,
and not only some subsystems.
And indeed, the projector ${\bf P}_{+}$ onto the globally symmetric space ${\cal H}\odot{\cal H}$
is given by the sum over all such terms.
In turn, the operator $A$ defined above in Eq.~(\ref{Projectors}) is -- up to the prefactor of $4$ --
the projector onto all globally symmetric states with the only exception of those states that are symmetric in every subsystem.
Thus, more formally, $A$ reads
\begin{equation}
A=4\left({\bf P}_{+}-P^{1}_{+}\otimes\hdots\otimes P^{N}_{+}\right)\ .
\end{equation}
Now, the two--fold copy $\ket{\Psi_{N}}\otimes\ket{\Psi_{N}}$ of an arbitrary pure state $\ket{\Psi_{N}}$
-- separable or not -- is always globally symmetric, {\it i.e.}
$\ket{\Psi_{N}}\otimes\ket{\Psi_{N}}\in{\cal H}\odot{\cal H}$.
On the other hand, any term
$P^{1}_{s_{1}}\otimes\hdots\otimes P^{N}_{s_{N}}$,
with an odd number of antisymmetric projectors, projects onto states that are globally antisymmetric,
{\it i.e.} onto a space that is orthogonal to ${\cal H}\odot{\cal H}$.
Therefore, the expectation value of such a term with respect to a two--fold copy
$\ket{\Psi_{N}}\otimes\ket{\Psi_{N}}$ always vanishes,
which is the reason to restrict the sum in Eq.~(\ref{Projectors}) to only terms with an even number of antisymmetric projectors.
Thus, one can add to $A$ any contribution of operators that are supported only on
${\cal H}\wedge{\cal H}$ without changing the value of $C_{N}$.
In this particular case, it turns out most useful to add the projector ${\bf P}_{-}$ onto the globally antisymmetric space
${\cal H}\wedge{\cal H}$, weighted with a prefactor $4$.
Since ${\bf P}_{-}$ and ${\bf P}_{+}$ add up to the identity ${\bm 1}$,
this amounts to replacing $A$ by $\tilde A$ in Eq. (\ref{Product}), being
\begin{equation}
\tilde A=4\left({\bm 1}-P^{1}_{+}\otimes\hdots\otimes P^{N}_{+}\right)\ .
\end{equation}
Thus, $C_{N}$ can be expressed in terms of one single {\em factorizable} observable,
which is in contrast to the  $2^{N-1}-1$ terms composing $A$, required to construct $C_{N}$ througth Eq.~(\ref{Projectors}).
Therefore, it can be experimentally determined through the measurment of only one single probability $p^{N}_{+}$ to find each of all $N$ subsystems and their copies
in a symmetric state, via
\begin{equation}
\label{Optimal}
C_{N}(\Psi_{N})=2\sqrt{1-p^{N}_{+}} \ .
\end{equation}
Finaly, it is even possible to reduce the number of subsystems on which to measure, as there  exists a redundancy in the $N$--partite measurement.
Since the two--fold copy $\ket{\Psi_{N}}\otimes\ket{\Psi_{N}}$ of a pure state is globally symmetric, it is indeed sufficient to determine the probability of $N-1$
subsystems and copies being in a symmetric state.
After the projection of the two--fold copies of any $N-1$ subsystems onto their symmetric subspaces, the remaining two--fold copy is
automatically projected onto its symmetric subspace as well.
Thus, the probability of finding all $N$ duplicate subsystems symmetric is equal to its analogous quantity for only $N-1$ subsystems, {\it i.e.}
$p^{N}_{+}=p^{N-1}_{+}$.
\par This redundancy turns out to be very useful to check whether the system is really in a pure state.
If one does observe a finite number of events where an odd number of subsystems is in an antisymmetric state,
then this indicates that the state in question is not pure, and needs to be described by a density matrix $\varrho$.
The probability of observing an antisymmetric state gives a quantification of the degree of mixing,
which can be expressed as
$1-\mbox{Tr}\varrho^{2}=2\mbox{Tr}({\bf P}_{-}\varrho\otimes\varrho)$.
Now, analogously to the case of ${\bf P}_{+}$ above,
the projector ${\bf P}_{-}$ onto the globally antisymmetric subspace decomposes into all products of $P_{-}^{i}$, and $P_{+}^{i}$ with an odd number of antisymmetric
factors.
Therefore, adding up the probabilities of observing the corresponding events, allows to experimentally determine the degree of mixing of $\varrho$ with exactly the same setup
used to measure the concurrence.
%%%%%%%%%%%%%%%%%%%%%%%%%%%%%%%%%%%%%%%%%%%%%%%%%%%%%%%%%%%%%%%%%%%%%%%%%%%%%%%%%%%%%%%%%%%%%%%%%%%%%%%%%%%%%%%%%%%%%%%%%%%%%%%%
\par \emph{Application to entangled--photon experiments}.--- Entangled photons supply us with a system to which the last result is
particularly relevant, for the single--setting--measurement scheme used in \cite{Steve} can be immediately extended to more than two
photons. In particular, the experiment that we have in mind is one in which  the techniques described in
\cite{Eibl,Witnesses, Bowmeester, Pan, Pan2, Zhao} to create $3$, $4$ or even $5$ entangled photons are combined with the hyperentanglement
techniques described in \cite{Kwiat} and used in \cite{Steve} to create copies of polarization states in the transverse momentum degrees of freedom, so that a multi--photon
entangled state is encoded into the photons polarizations and the copy in the momenta. If such a state is realized with $N$ photons,
it is only necessary to perform the two--qubit single--photon Bell--state measurement described in \cite{Steve} on the polarization--momentum states carried
by any $N-1$ of the $N$ photons, to obtain the probability
$p^{N-1}_{ +}=p^{N}_{+}$ of every polarization qubit and its momentum--qubit copy being in a symmetrical state. Alternatively, other
photon spatial degrees of freedom can be used to encode the copy as well, as for example the first order Hermite--Gaussian modes, for which unambiguous
perfect--efficiency single--photon Bell--state analyzers have also been constructed with linear optical devices \cite{Fiorentino-Steve}.

\par Finally, we emphasize that all
measurements are performed locally, as the two qubits are always encoded in the same photon; that the detection in the Bell basis can be done by
linear--optics Bell--state analyzers; and that discrimination among all four bell states is not required, but rather only between the antisymmetric singlet
and the remaining symmetric Bell states. An implementation of our single--setting detection strategy with four or five entangled photons
with already existing technology thus seems feasible.
%%%%%%%%%%%%%%%%%%%%%%%%%%%%%%%%%%%%%%%%%%%%%%%%%%%%%%%%%%%%%%%%%%%%%%%%%%%%%%%%%%%%%%%%%%%%%%%%%%%%%%%%%%%%%%%%%%%%%%%%%%%%%%%%
\par \emph{Application to trapped--ion experiments}.--- Trapped--ions provide another system in which the state--of--the--art of technology allows for an immediate
implementation of the scheme developed here. Let us consider an experiment with $2N$  ions trapped in a linear Paul trap with individual laser
addressing to each ion. In a first stage of the experiment a GHZ or W state is created in $N$ ions using the techniques described in \cite{Roos, Sackett, Leibfried} or
\cite{Haeffner}, respectively.
In this stage, a collective motional mode is used as the ``information--bus'' among the different ions on which the laser beams shine
and is brought to its same initial state in the end. Also, as no laser beam shines on the other $N$ ions, their internal states remain
untouched. The result of this first stage is an entangled state encoded into the internal
state of the first $N$ ions, and the initial state untouched for the $2N$--ion collective motional mode and internal modes of the
second $N$ ions.

\par Once the entangled state is created in the first $N$ ions, the
same procedure is used on the second set of ions to create the copy. %In this second stage it is the first $N$ ions that are now not manipulated by any laser field
%and are thus left appart.
After these two stages, the resulting $2N$--ion state is one in which the first $N$ ions share an entangled state and the second $N$
ions share the copy. Then, it is just a matter of choosing any $N-1$ out of the $N$ first ions and measuring each one, with its copy in the
second set, in the Bell basis to obtain the probability  $p^{N-1}_{+}$ and thus calculate $C_{N}$ using Eq. (\ref{Optimal}).
The Bell--state detection, in turn, can be performed by running the sequence of pulses used in \cite{Roos2} backwards, in which all
four Bell states were created starting from the product states of the computational basis. In this way, each Bell state can be mapped into a different product state of
the  computational basis and then finally measured with usual (almost--unit--efficiency) state--selective fluorescence detection
\cite{Flourescence, Roos2}.
%%%%%%%%%%%%%%%%%%%%%%%%%%%%%%%%%%%%%%%%%%%%%%%%%%%%%%%%%%%%%%%%%%%%%%%%%%%%%%%%%%%%%%%%%%%%%%%%%%%%%%%%%%%%%%%%%%%%%%%%%%%%%%%%
\par
\emph{Summary}.---
We showed that for a composite system, with any number of arbitrary--finite--dimensional subsystems, for which a copy of its state is available,
it is possible to express the multipartite concurrence of pure states in terms of only one single factorizable observable. This result has immediate
utility on trapped--ion and entangled--photon experiments, for which we showed how a direct measurement, with a single experimental setting, of the multipartite
concurrence of pure states of photons and ions is feasible with the use of already existing technology.
%%%%%%%%%%%%%%%%%%%%%%%%%%%%%%%%%%%%%%%%%%%%%%%%%%%%%%%%%%%%%%%%%%%%%%%%%%%%%%%%%%%%%%%%%%%%%%%%%%%%%%%%%%%%%%%%%%%%%%%%%%%%%%%%
\par \emph{Acknowledgements}.--- We would like to thank Alejo
Salles and Andreas Buchleitner  for helpful discussions; and Stephen Walborn for careful reading of the manuscript.
\par This work was financially supported by the CNPq, the FAPERJ, the CAPES and the Brazilian Millenium Institute for Quantum Information.
F.M. gratefully acknowledges financial support by the Alexander von Humboldt Foundation.
%%%%%%%%%%%%%%%%%%%%%%%%%%%%%%%%%%%%%%%%%%%%%%%%%%%%%%%%%%%%%%%%%%%%%%%%%%%%%%%%%%%%%%%%%%%%%%%%%%%%%%%%%%%%%%%%%%%%%%%%%%%%%%%%

%%%%%%%%%%%%%%%%%%%%%%%%%%%%%%%%%%%%%%%%%%%%%%%%%%%%%%%%%%%%%%%%%%%%%%%%%%%%%%%%%%%%%%%%%%%%%%%%%%%%%%%%%%%%%%%%%%%%%%%%%%%%%%%%
%%%%%%%%%%%%%%%%%%%%%%%%%%%%%%%%%%%%%%%%%%%%%%%%%%%%%%%%%%%%%%%%%%%%%%%%%%%%%%%%%%%%%%%%%%%%%%%%%%%%%%%%%%%%%%%%%%%%%%%%%%%%%%%%

\begin{thebibliography}{99}
\bibitem{Plenio}
M. Horodecki, Quant. Inf. Comp. 1,
3 (2001); P. Horodecki, R. Horodecki, Quant. Inf. Comp. 1, 45 (2001); J. Eisert,
M. B. Plenio, Int. J. Quant. Inf. 1, 479 (2003); M. B. Plenio, S. Virmani, % {\em An introduction to entanglement measures},
E-print arxiv quant-ph/0504163.
\bibitem{Hills-Wooters}
S. Hills, W. K. Wootters, Phys. Rev. Lett. {\bf 78}, 5022 (1997).
\bibitem{Wooters}
W. K. Wootters, %{\em Entanglement of formation of an arbitrary state of two qubits},
Phys. Rev. Lett. {\bf 80}, 2245 (1998).
\bibitem{Rungta}
P. Rungta, V. Buzek, C. M.
Caves, M. Hillery, G. J. Milburn, %{\em Universal state inversion  and concurrence in arbitrary dimensions},
Phys. Rev. A {\bf 64}, 042315 (2001).
\bibitem{Ulhmann}
 A. Ulhmann, %{\em Fidelity and concurrence of conjugated states},
Phys. Rev. A {\bf 62}, 032307 (2000).
\bibitem{Meyer-Wallach-Brennen}
D. A. Meyer, N. R. Wallach, J. Math. Phys. (N. Y.) {\bf 43}, 4273 (2002); G. K. Brennen, Quantum Inf. Comput. {\bf 3}, 619 (2003).
\bibitem{Andre}
A. R. R. Carvalho, F. Mintert, A. Buchleitner, %{\em Decoherence and multipartite entanglement},
Phys. Rev. Lett {\bf 93}, 230501 (2004).
\bibitem{Flo}
F. Mintert, M. Ku\'{s}, A. Buchleitner, %{\em Concurrence of mixed bipartite quantum states in arbitrary dimensions},
Phys. Rev. Lett. {\bf 93}, 230501 (2004)
\bibitem{Flo-Review}
F. Mintert, M. Ku\'{s}, A. Buchleitner, Phys. Rev. Lett. 95, 260502 (2005); F. Mintert, A. R. R. Carvalho, M. Ku\'{s}, A. Buchleitner, %{\em Measures and dynamics of entangled states},
Phys. Rep. {\bf 415}, 207 (2005).
\bibitem{Steve}
S. P. Walborn, P. H. Souto, L. Davidovich, F. Mintert, A. Buchleitner, %{\em Single-shot measurement of an entanglement measure},
Nature 440, 1022 (2006) .
\bibitem{Eibl}
M. Eibl, N. Kiesel, M. Bourennane, C. Kurtsiefer, H. Weinfurter, %{\em Experimental realization of a three-qubit entangled W-state},
Phys. Rev. Lett. {\bf 92}, 077901 (2004).
\bibitem{Witnesses}
M. Bourennane, M. Eibl, C. Kurtsiefer, S. Gaertner, H. Weinfurter, O. G\"{u}hne, P. Hyllus, D. Bruss, M. Lewenstein, and A. Sanpera,
Phys. Rev. Lett. 92, 087902 (2004).
\bibitem{Bowmeester}
D. Bowmeester, J.-W. Pan,  M. Daniell, H Weinfurter, A. Zeilinger, %{\em Experimental demonstration of four-photon entanglement and high-fidelity teleportation}
Phys. Rev. lett. {\bf 82}, 1345 (1999).
\bibitem{Pan}
J.-W. Pan, D. Bowmeester, M. Daniell, H Weinfurter, A. Zeilinger, %{\em Observation of three-photon Greenberger-Horn-Zeilinger entanglement},
Nature {\bf 403}, 515 (2000).
\bibitem{Pan2}
J.-W. Pan, M. Daniel, S. Gasparoni, G. Weihs, A. Zeilinger, %{\em Experimental test of quantum non-locality in three-photon Greenberger-Horn-Zeilinger-entanglement},
Phys. Rev. Lett. {\bf 86}, 4435 (2001).
\bibitem{Zhao}
Z. Zhao, Y.-A. Chen, A.-N Zhang, T. Yang, H. J. Briegel, J.-W. Pan, %{\em Experimental demonstration of five-photon entanglement and open-destination teleportation},
Nature {\bf 430}, 54 (2004).
\bibitem{Raschenbeutel}
A. Raschenbeutel, G. Nogues, S. Osnaghi, P. Bertet, M. Brune, J.-M. Raimond, S. Haroche, %{\em Step-by-step engeneered multiparticle entanglement},
Science {\bf 288}, 2024 (2000).
\bibitem{Roos}
C. F. Roos, M. Riebe, H. H\"{a}ffner, W. H\"{a}nsel, J. B. Elm, G. P. T. Lancaster, C. Becher, F. Schmidt-Kaler, R. Blatt, %{\em Control and Measurement of three-qubit entangled states},
Science {\bf 304}, 1478 (2004).
\bibitem{Sackett}
C. A. Sackett, D. Kielpinski, B. E. King, C.
Langer, V. Meyer, C. J. Myatt, M. Rowe, Q. A. Turchette, W. M.
Itano, E. J. Wineland, and C. Monroe, %{\em Experimental entanglement of four particles},
Nature \bf 404\rm, 256 (2000).
\bibitem{Leibfried}
D. Leibfried, E. Knill, S. Seidelin, J. Britton, R. B. Blakestad, J. Chiaverini, D. B. Hume, W. M. Itano, J. D. Jost, C. Langer, R. Ozeri, R. Reichle, D. J. Wineland,
Nature {\bf 438}, 639 (2005).
\bibitem{Haeffner}
H. H\"{a}ffner, W. H\"{a}nsel, C. F. Roos, J. Benhelm, D. Chek-al-kar, M. Chwalla, T. Kšrber, U. D. Rapol, M. Riebe, P. O. Schmidt, C. Becher, O. G\"{u}hne, W. D\"{u}r, R. Blatt, %{\em Scalable multiparticle entanglement of trapped ions},
Nature {\bf 438}, 643 (2005).
\bibitem{White}
A. G. White, D. F. V. James, P. H. Eberhard, P. G. Kwiat, Phys. Rev. Lett. {\bf 83}, 3103 (1999).
\bibitem{Roos2}
C. F. Roos, G. P. T. Lancaster, M. Riebe, H. H\"{a}ffner, W. H\"{a}nsel, S. Gulde, C. Becher, J. Eschner, F. Schmidt-Kaler, R. Blatt, Phys. Rev. Lett. {\bf 92}, 220402 (2004).
\bibitem{HoroWitness}
M. Horodecki, P. Horodecki, R. Horodecki, Phys. Lett. A {\bf 223}, 1 (1996).
\bibitem{Kwiat}
P. G. Kwiat, J. Mod. Opt.{\bf  44}, 2173 (1997); J. T. Barreiro, N. K. Langford, N. A. Peters, and P. G. Kwiat, %Generation of Hyperentangled Photon Pairs
Phys. Rev. Lett. {\bf 95}, 260501 (2005).
\bibitem{Fiorentino-Steve}
M. Fiorentino, F. N. C. Wong, Phys. Rev. Lett. {\bf  93}, 070502 (2004); A. N. de Oliveira,  S. P. Walborn, C. H. Monken, Journal of Optics B: Quantum and Semiclassical
Optics {\bf  7}, 288 (2005).
\bibitem{Flourescence}
F. Schmidt-Kaler, H. H\"{a}ffner, S. Gulde, M. Riebe, G.P.T. Lancaster, T. Deuschle, C. Becher, W. H\"{a}nsel, J. Eschner, C.F. Roos and R.
Blatt, Appl. Phys. B {\bf  77}, 789 (2003).
\end{thebibliography}
\end{document}